\documentclass[twocolumn,nofootinbib,aps,prd,superscriptaddress, 10pt]{revtex4-1}
\usepackage{amsmath,graphicx,multirow,hyperref,url,color,rotating,amssymb,bm}

\begin{document}

\title{A User-Friendly Dark Energy Model Generator}

\author{Kyle A. Hinton}
\email{kahinton@umich.edu}
\affiliation{Department of Physics, University of Michigan, 450 Church St, Ann Arbor, MI 48109-1040}

\author{Adam Becker}
\email{adam@freelanceastro.com}
\affiliation{Freelance, Oakland, CA 94610}

\author{Dragan Huterer}
\email{huterer@umich.edu}
\affiliation{Department of Physics, University of Michigan, 450 Church St, Ann
  Arbor, MI 48109-1040}
\affiliation{Max-Planck-Institut for Astrophysics, Karl-Schwarzschild-Str.\ 1, 85741 Garching, Germany}
\affiliation{Excellence Cluster Universe, Technische Universit\"{a}t M\"{u}nchen, Boltzmannstrasse 2, 85748 Garching,
Germany}

\begin{abstract}
We provide software with a graphical user interface to calculate the
phenomenology of a wide class of dark energy models featuring multiple scalar
fields and potentials that are arbitrary functions of exponentials. The user
chooses a subclass of models and, if desired, initial conditions, or else a
range of initial parameters for Monte Carlo. The code calculates the energy
density of components in the universe, the equation of state of dark energy,
and the linear growth of density perturbations, all as a function of redshift
and scale factor. The output also includes an approximate conversion into the
average equation of state, as well as the common $(w_0, w_a)$
parametrization. The code is available here:
\url{http://github.com/kahinton/Dark-Energy-UI-and-MC}
\end{abstract}

\maketitle

\section{Introduction}\label{sec:intro}

The puzzle of dark energy is one of the most important outstanding questions
in all of physics. Studying the observational signatures of dark energy -- and
using those signatures to understand the nature of dark energy -- has been a
highly active area of research since the first signs of dark energy were
discovered over 15 years ago. (For a review, see
e.g.\ \citet{Frieman:2008sn}.) Dark energy models have a rich phenomenology,
leaving varied fingerprints on the growth of cosmic structure and on
geometrical quantities in the universe.

We have written a piece of software that makes it easier to compute
quantitative predictions for dark energy models broader than a simple
cosmological constant. These models possess a richer phenomenology that can be
tested with ongoing and future experiments. They also complement the
  choice of dark energy models -- and computer code -- made by others
  \cite{Marsh:2014xoa}. Our software is capable of handling a wide class of
models with multiple scalar fields, and specifically makes two subclasses
particularly easy to calculate with a simple user input. In these models, the
user can calculate the expansion history and the growth of structure in the
linear regime, therefore enabling the user to obtain various implications for
observable quantities in cosmology. The code also returns an approximate
conversion of the equation of state of dark energy $w(a)$ into two commonly
used parameters $(w_0, w_a)$, as well as several other commonly used
parameters.

\begin{figure}[b]
\includegraphics[width=0.45\textwidth]{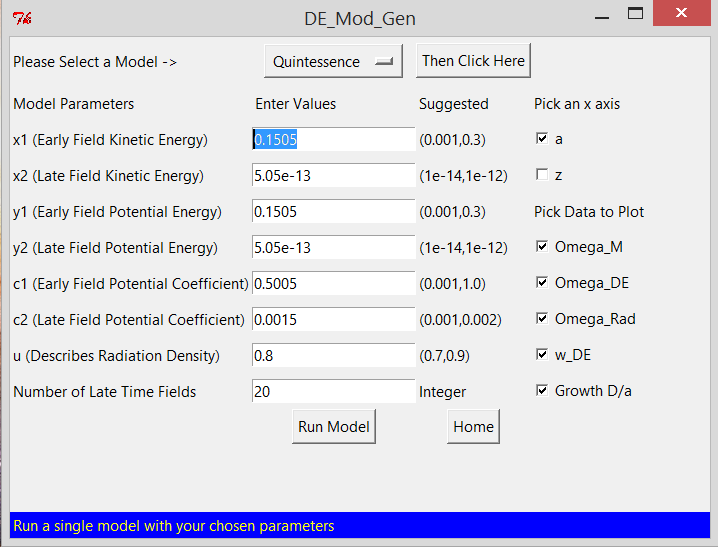}
\caption{Sample screenshot of the software, showing text inputs for dark
  energy model parameters, as well as checkboxes for selecting the desired
  plots.}
\label{fig:screen}
\end{figure}

\begin{figure*}[t]
\includegraphics[width=0.49\textwidth]{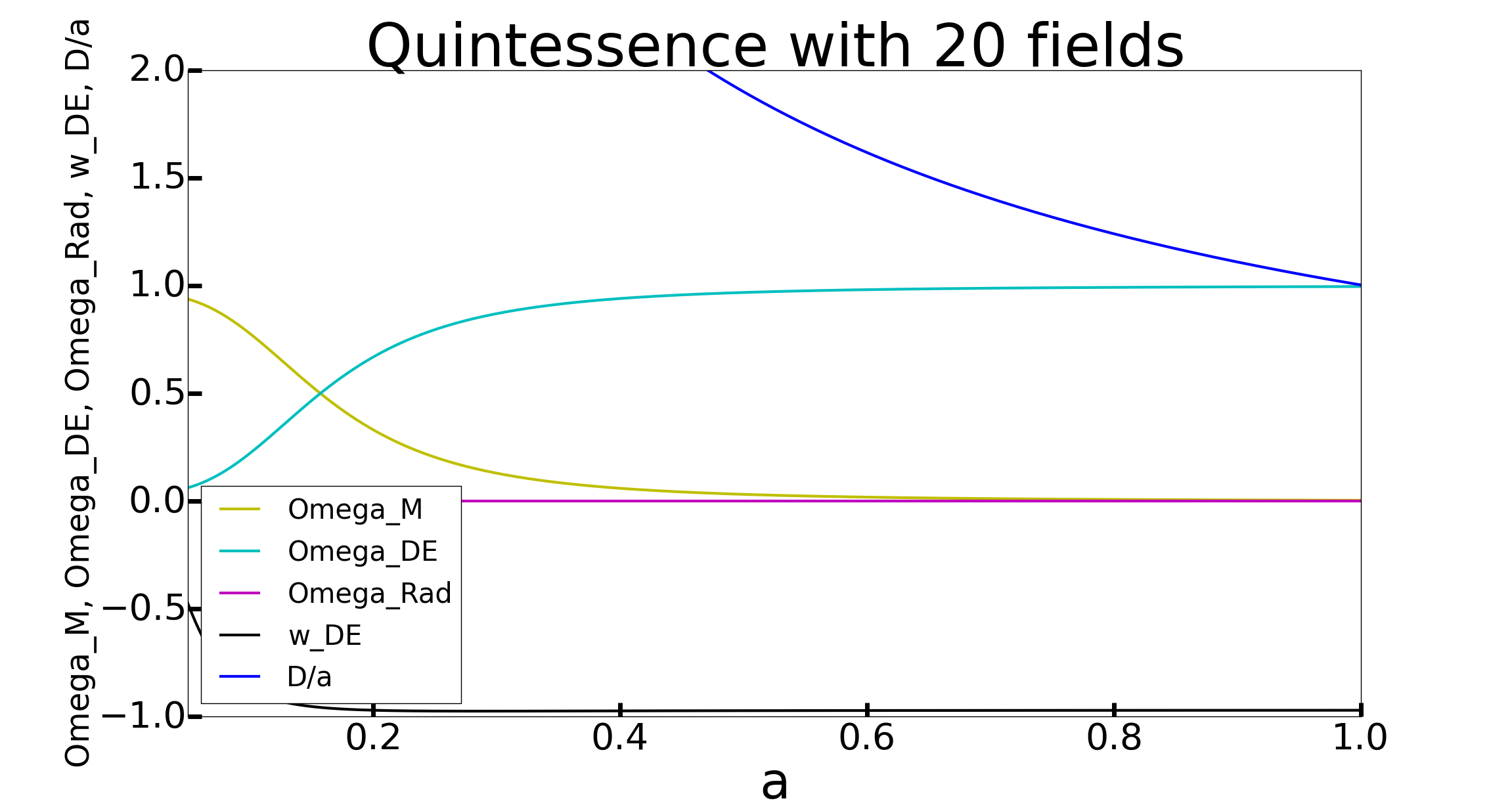}
\includegraphics[width=0.49\textwidth]{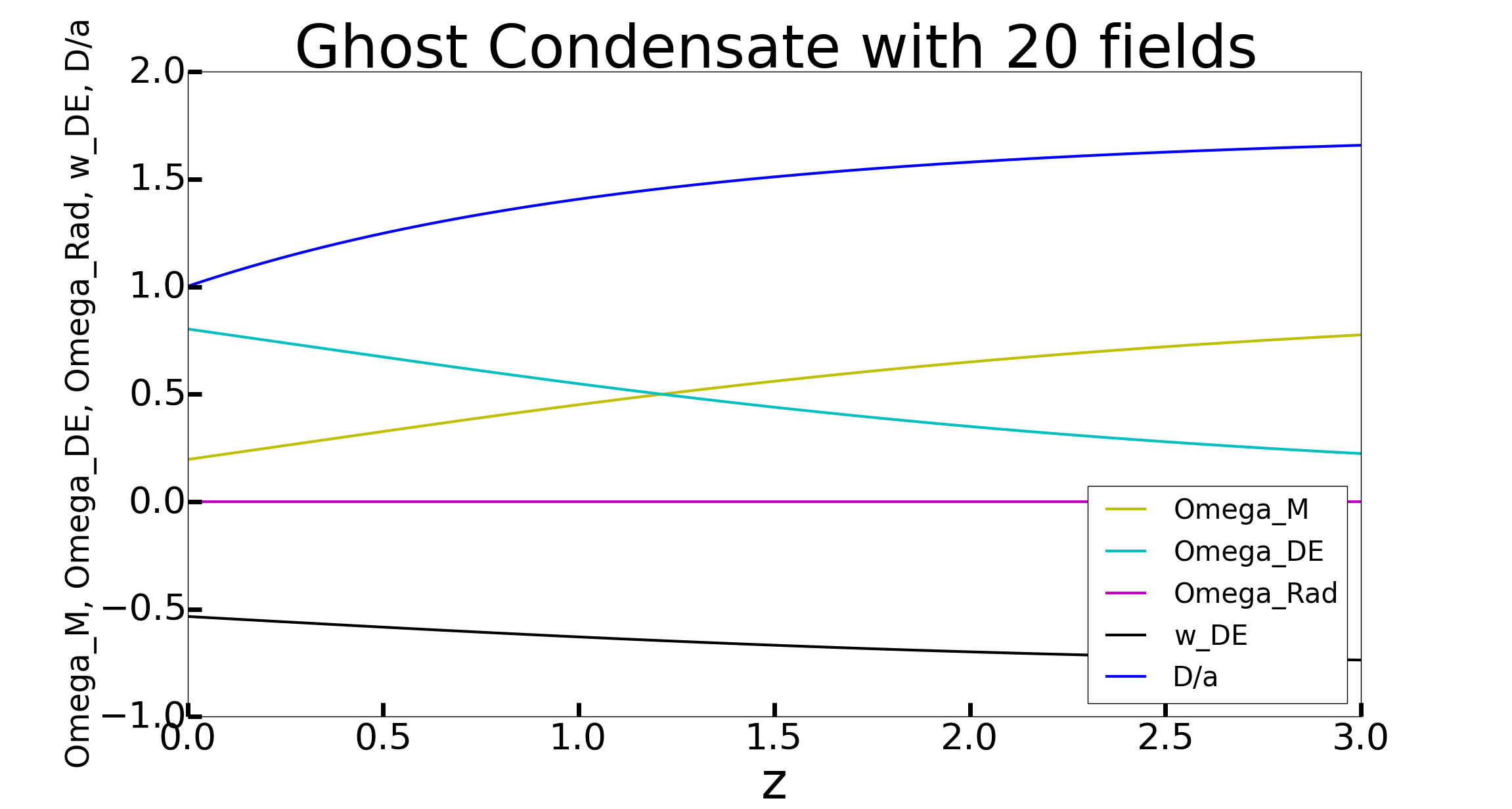}
\caption{Example plots generated by our program, showing output for a single
  model for our quintessence (left panel) or ghost condensate (right panel)
  class of models. The user can choose to plot this for any arbitrary choice of
  model parameters, as a function of either scale factor $a$ or redshift $z$.}
\label{fig:omegas_vs_a}
\end{figure*}

\section{Dark Energy Models}

We choose two classes of ``assisted dark energy'' models, based specifically
on the analysis  by  \citet{Ohashi:2009xw}, and originally developed by
Refs.~\cite{Piazza:2004df,Tsujikawa:2004dp}.  The starting assumption is the
existence of $n$ scalar fields, and a Lagrangian density that takes the
following form
\begin{equation}
\mathcal{L}_{\rm fields} = \sum_{i=1}^n X_i g(X_ie^{\lambda_i\phi_i})
\end{equation}
where $X_i\equiv -g^{\mu\nu}\partial_\mu \phi_i\partial_\nu \phi_i /2$ is the
kinetic term for $i$-th field, and $g(Y)$ is an arbitrary function of its
argument $Y\equiv X\,e^{\lambda\phi}$, where $\lambda$ is a dimensionless
parameter. The existence of scaling solutions dictates that this be the most
general form of the Lagrangian density for exponential potentials
\cite{Piazza:2004df,Tsujikawa:2004dp}. A noteworthy special case of our class
of models is quintessence with a single field ($n=1$), and a simple
exponential potential $V(\phi)=c\exp(\lambda\phi)$, for which $g(Y)=1-c/Y$.
As we explain below, this class of models also includes ghost
condensate models \cite{ArkaniHamed:2003uy} featuring a background scalar
field which, despite having a kinetic term of the ``wrong'' sign, are stable
and feasible due to suitable higher-order kinetic terms.

Our code allows a user to test models with an arbitrary function $g(Y)$, for an
arbitrary number of fields $n$. As an example we also provide two special cases
which have been run through the Monte Carlo generator, and whose results can
be accessed with very simple inputs in the graphical user interface (GUI):
\begin{eqnarray}
\label{eq:gY_quint}
g(Y_i)  &=& 1-c_i/Y_i\quad \mbox{(quintessence})\\[0.2cm]
g(Y_i)  &=& -1+c_iY_i\quad \mbox{(ghost condensate}).
\label{eq:gY_ghost}
\end{eqnarray}
In each case, the temporal evolution of all quantities is most easily tracked
in terms of scaled variables
\begin{equation}
x_i\equiv \frac{\dot{\phi}_i}{\sqrt{6}H};\quad
y_i\equiv \frac{e^{-\lambda_i\phi_i/2}}{\sqrt{3}H};\quad
u\equiv \frac{\sqrt{\rho_r}}{\sqrt{3}H}
\end{equation}
where $\rho_r$ is the radiation energy density. The evolution equations,
expressed in terms of the time variable $N\equiv \ln(a)$, are \citep{Ohashi:2009xw}
\begin{eqnarray}
\frac{d x_i}{d N}&=& \frac{x_i}{2} \left[ 3+3\sum_{i=1}^n
x_i^2 g(Y_i)+u^2-\sqrt{6}\lambda_i x_i \right] 
\nonumber \\
&&+\frac{\sqrt{6}}{2}A(Y_i) \left[ \lambda_i \Omega_{\phi_i}
-\sqrt{6} \{ g(Y_i)+Y_ig'(Y_i) \}x_i \right]\,, 
\nonumber \\[0.2cm]
\frac{dy_i}{dN} &=& \frac{y_i}{2} \left[ 3+3\sum_{i=1}^n
x_i^2 g(Y_i)+u^2-\sqrt{6}\lambda_ix_i \right]\,,
\label{auto2} \\[0.2cm]
\frac{du}{d N}  &=& \frac{u}{2} \left[-1+3\sum_{i=1}^n
x_i^2 g(Y_i)+u^2 \right].
\label{auto3}
\end{eqnarray}

We numerically evolve these equations. Then all physical quantities of interest 
can be found, for example the energy densities of matter and dark energy
\begin{equation}
\Omega_{\rm DE} = \sum_{i=1}^n x_i^2 \left[ g(Y_i)+2Y_ig'(Y_i) \right]
\label{Omega_phi}
\end{equation}
and the dark energy equation of state
\begin{equation}
w_{\rm DE}\equiv \frac{\sum_{i=1}^n p_{\phi_i}}
{\sum_{i=1}^n \rho_{\phi_i}}=
\frac{\sum_{i=1}^n x_i^2 g(Y_i)}
{\sum_{i=1}^n x_i^2 [g(Y_i)+2Y_ig'(Y_i)]}
\end{equation}
where the time-dependent parameter $Y_i$ can be calculated via
$Y_i = x_i^2/y_i^2$.
Note that for quintessence, the functional form of $g(Y_i)$ in
Eq.~(\ref{eq:gY_quint}) implies that the potential is the sum of exponential
potentials for each field, $V(\phi)=\sum_i c_i \exp(\lambda_i\phi_i)$.

\section{Using the Provided Software}

Our code can be found at the following GitHub repository:
\url{http://github.com/kahinton/Dark-Energy-UI-and-MC}.  The code gives the
user several tools for testing and analyzing various models of dark
energy. The first step in being able to use these tools is to run a model
through our Monte Carlo generator.  The user must supply several inputs to the
MC generator: a specific function $g(Y)$, as described above; a name to
differentiate the model from others; and the number of different initial
conditions to test the model over. The provided name will be used to label
this model in the user interface. The user must also specify the values for
several parameters governing each model. Specifically these parameters are the
number of fields that will be acting as dark energy, $n$, as well as the range
for the initial conditions of $x_i$, $y_i$, $c_i$, and $u$. In our code we
adopt a simplification suggested in \citep{Ohashi:2009xw}, which assumes that
a single field will act to assist inflation in the early universe before
becoming small, while the remaining fields will be small in the early universe
before becoming the dominant form of energy at late times. The ``assisted
inflation'' field will have one set of unique initial conditions, while the
remaining $n-1$ fields will all share the same initial conditions. This later
simplification drives the value of $w_{DE}$ as close to $-1$ as possible at
late times, and also allows models to be tested at the same speed independent
of the number of fields being used.

\begin{figure*}[t]
\includegraphics[width=0.40\textwidth]{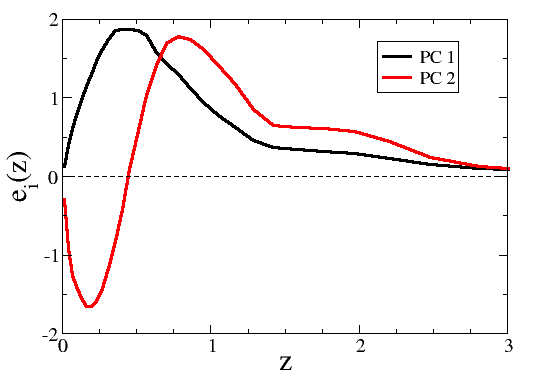}
\includegraphics[width=0.52\textwidth]{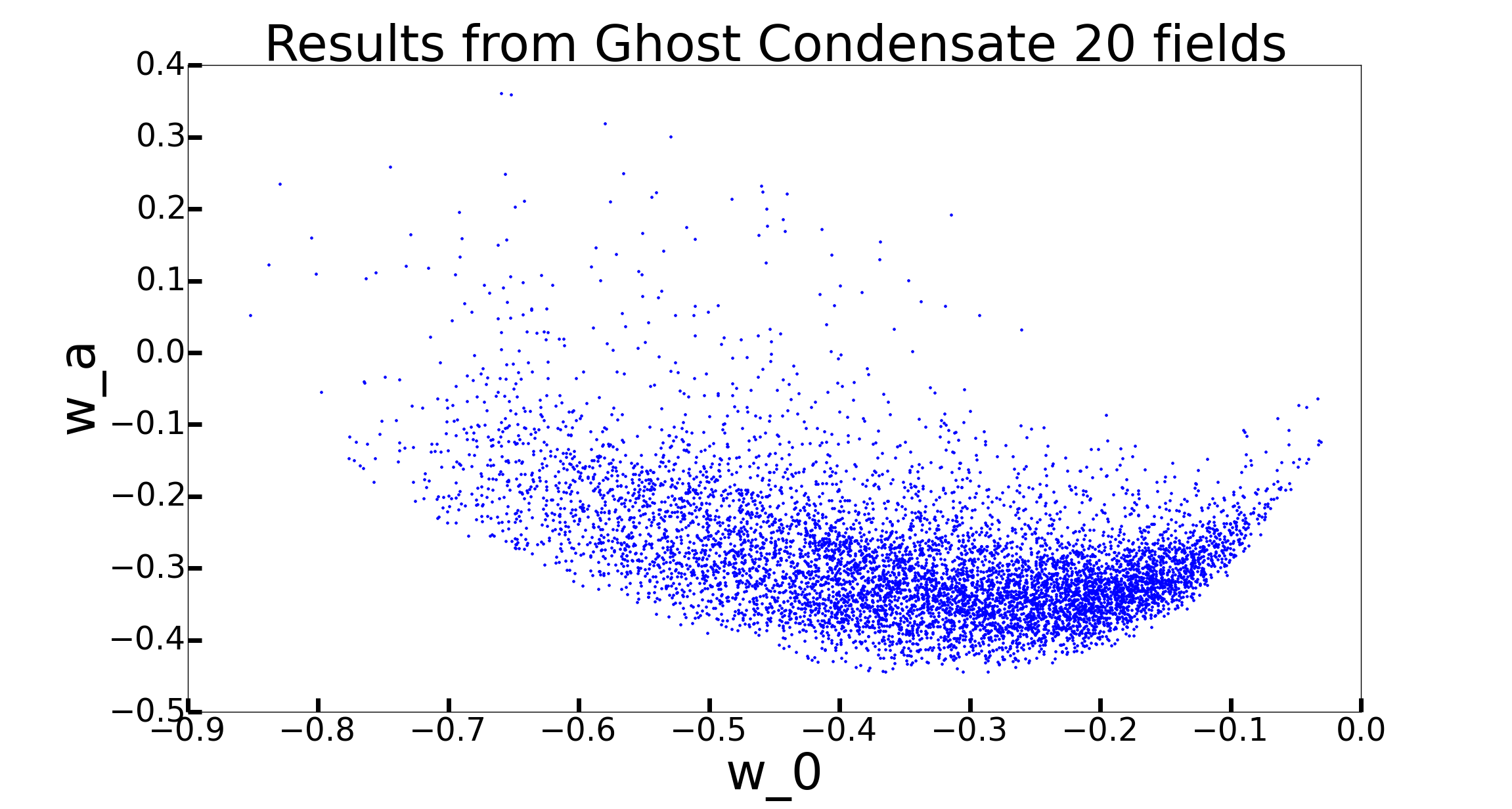}
\caption{Left panel: the first two principal components used to convert from
  $w(z)$ to $w_{\rm ave}$ and $(w_0, w_a)$; note that $w_{\rm ave}$ requires only
  the first principal component. The components are based on the NASA/DOE Figure of
  Merit Science Working Group analysis \cite{albrecht:2009ct}; see Appendix
  for details. Right panel: a sample scatterplot made from the user
  interface. Here one can see the results of $w_0$ and $w_a$ for 10,000
  pre-supplied ghost condensate models, case with 15 fields.}
\label{fig:w_plot}
\end{figure*}

By using these simplifications the user only needs to specify the range of
initial conditions for the first two fields. The code then assumes, for
example, that all values of $x_2$ to $x_n$ are equal. To ensure that a scalar
field dominated solution is produced, the values for $\lambda_i$ are chosen
automatically, according to the inequalities (see Eqs.~(29) and (32) in
\citep{Ohashi:2009xw})
 
\begin{equation}
\frac{\lambda_{\rm 1}^2}{p_{,X}} > 88.9;\quad
\frac{\lambda_{\rm 2-n}^2}{p_{,X}} < 2,
\end{equation}
Here $p$ is the Lagrangian density, and
$p_{,X}\equiv\partial{p}/\partial{X}$. When the code is run, for each specific
test, the initial conditions for each parameter are chosen from a random flat
distribution over the selected range. When all of the tests have finished
running the model will be added to the user interface automatically. As an
example of what is output to the user interface, we have included the results
of tests of quintessence and ghost condensate models having 5, 10, 15, and 20
fields.

The user interface also contains two optional tools. The first of these allows
the user to run a single instance of any model having been run through the
Monte Carlo generator. Each of the required parameters are assigned a set of
fiducial values that depend on which model is chosen; however, the range over
which the model has been tested is also supplied, allowing the user to
experiment with values other than those provided. Once the user selects values
for these parameters, they can choose to plot various quantities as a function
of either scale factor $a$ or redshift $z$. These quantities include:


\begin{itemize}
\item Energy densities in units of critical (radiation $\Omega_R(a)$, matter
  $\Omega_M(a)$, dark energy $\Omega_{\rm DE}(a)$);
\item Equation of state of dark energy $w(a)$;
\item Linear growth of density fluctuations $D(a)$
 (we plot the quantity $D(a)/a$).
\end{itemize} 

\noindent In order to provide the growth of linear perturbations,
concurrently with the other equations we also evolve
\begin{equation}
D''+D'(H'/H+2)-3/2(H_0/H)^2\Omega_M(1+z)^3D=0
\end{equation}
where $D(a)\equiv (\delta\rho/\rho) / (\delta\rho/\rho)_{a=1}$ is the linear
growth rate of matter perturbations, and where the primes indicate a derivative with
respect to $\ln(a)$.  The quantity $D/a$ for the selected model will
immediately be displayed on screen for the user to examine.

The second option available to the user is to examine the results from the
Monte Carlo generator. Specifically, this tool allows the user to generate
two-dimensional scatterplots on the fly for further analysis -- for example,
to see what range of phenomenological outcomes is produced by a given model or
class or models. Each point in a given scatterplot corresponds to an output of
a single model. The output can be any of the following parameters or functions:

\begin{itemize}
\item Initial conditions from each model including $x_1$, $x_2$, $y_1$, $y_2$,
$c_1$, $c_2$, $\lambda_1$, $\lambda_2$, $u$, and $n$;
\item Energy densities in matter and dark energy today $\Omega_M$ and
  $\Omega_{\rm DE}$, as well as the equation of state today $w_{\rm DE}(a=1)$.
\item Effective ``average'' equation of state $w_{\rm ave}$, calculated from
  the first principal component, as described in the Appendix.
\item Effective values for the parameters $w_0$ and $w_a$ based on
  parametrization \citep{Linder:2002et} $w(a)=w_0+w_a(1-a)$, derived from the
  first two principal components as described in the Appendix.
\end{itemize}
An example of the Monte Carlo output showing the $w_0-w_a$ plane 
is shown in the right panel of Fig.~\ref{fig:w_plot}.

\section*{Acknowledgments}
We thank Eric Linder for useful comments. 
Our work has been supported by NSF  under contract AST-0807564 and DOE under contract
DE-FG02-95ER40899. 

\appendix
\setcounter{secnumdepth}{0} 
\section{Appendix: Principal components, $w_{\rm ave}$, and ($w_0, w_a$)}\label{app}

Here we explain in more detail how to calculate the aforementioned quantities
$w_0$, $w_a$, and $w_p$ from the principal components of the equation of
state, following \cite{Huterer:2002hy,Huterer:2006mv}. First, we formally expand the equation
of state in terms of principal components 
\begin{equation}
1+w(a)=\sum_i \alpha_i e_i(a),
\end{equation} 
where $\alpha_i$ is the coefficient of the $i$-th principal component
$e_i(a)$. The idea is to convert the first principal component
$\alpha_1$ into the averaged value of the equation of state $w_{\rm ave}$, and
first {\it two} principal components $(\alpha_1, \alpha_2)$ into $(w_0, w_a)$.
This approach is justified because the first few principal components carry
essentially all the necessary information about the effects of dark energy
dynamics on the expansion of the universe on observable scales.

The principal components $e_i$ can be determined for any given dataset; here
we use the Figure of Merit Science Working Group's (FoMSWG) publicly available code
\cite{albrecht:2009ct}, and a combination of Planck+BAO+SN+WL available in
their code. Note that, while the component shapes --- especially the
all-important peak value of $e_1(a)$ that shows the temporal epoch of maximum
sensitivity to the equation of state --- depend on the cosmological probe as
well as specifications of a given experiment, once we combine the probes the
pull of different probes is expected to average out, leading to a fixed set of
shapes. 

While the 
normalization of the $e_i(a)$ is arbitrary in principle, the FoMSWG principal components that
we use are normalized as $\int e_i^2(a)da=1$. The coefficients $\alpha_i$ can be
obtained as
\begin{equation}
\alpha_i = \int [1+w(a)]\, e_i(a)\, da
\label{eq:alpha_i}
\end{equation}
where $w(a)$ is the actual equation of state of the theoretical dark energy
model we are studying. 

The final step is converting the first principal component into $w_{\rm ave}$,
and the first two into $w_0$ and $w_a$; we do this via
\citep{Huterer:2006mv}
\begin{equation}
1+w_{\rm ave} = \frac{\alpha_1}{\beta_1}\qquad
\mbox{(average $w$)}
\label{eq:w_ave}
\end{equation}
and 
\begin{equation}
\begin{split}
1+w_0 &\equiv  \frac{\alpha_1(\gamma_2 - \beta_2)+\alpha_2(\beta_1-\gamma_1)}
{\beta_1\gamma_2-\beta_2\gamma_1}\\[0.2cm]
w_a &\equiv  \frac{\alpha_1\beta_2-\alpha_2\beta_1}{\beta_1\gamma_2-\beta_2\gamma_1}
\label{eq:w0wa_from_alpha}
\end{split}
\end{equation}
where $\alpha_i$ are defined in Eq.~(\ref{eq:alpha_i}), and 
\begin{equation}
\beta_i \equiv \int e_i(a)\, da;\quad
\gamma_i\equiv \int a\,e_i(a)\, da
\end{equation} 

Equations (\ref{eq:w_ave}) and (\ref{eq:w0wa_from_alpha}) are now our
definitions of the parameters $w_{\rm ave}$ and $(w_0, w_a)$, respectively,
given a dark energy history $w(z)$ which determines the $\alpha_i$.  Previous work \citep{Huterer:2006mv} confirms that the two-parameter equation
of state closely follows the true $w(z)$ over the redshift range most
effectively probed by the data.

Note too that the constraint $w_0\geq -1$, which follows from $w(z)\geq -1$,
is not strictly obeyed by $w_0$ obtained in this way since $w_0$ and $w_a$ are
now essentially a {\it fit} to the dark energy equation of state history.  In
other words, the derived parameter $w_0$ {\it can} be slightly smaller than
$-1$; this is again because we are fitting these three $w$-parameters and
forcing them to the model's actual equation of state evolution. These
parameters are nevertheless provided since 1) they provide a very useful test
bed to gauge the effectiveness of the range of dark energy model behavior, and
2) they are fairly representative of these models, since actual dark energy models often do exhibit the
$w_0+w_a(1-a)$ scaling.

\bibliography{refs}

\begin{thebibliography}{10}%
\makeatletter
\providecommand \@ifxundefined [1]{%
 \@ifx{#1\undefined}
}%
\providecommand \@ifnum [1]{%
 \ifnum #1\expandafter \@firstoftwo
 \else \expandafter \@secondoftwo
 \fi
}%
\providecommand \@ifx [1]{%
 \ifx #1\expandafter \@firstoftwo
 \else \expandafter \@secondoftwo
 \fi
}%
\providecommand \natexlab [1]{#1}%
\providecommand \enquote  [1]{``#1''}%
\providecommand \bibnamefont  [1]{#1}%
\providecommand \bibfnamefont [1]{#1}%
\providecommand \citenamefont [1]{#1}%
\providecommand \href@noop [0]{\@secondoftwo}%
\providecommand \href [0]{\begingroup \@sanitize@url \@href}%
\providecommand \@href[1]{\@@startlink{#1}\@@href}%
\providecommand \@@href[1]{\endgroup#1\@@endlink}%
\providecommand \@sanitize@url [0]{\catcode `\\12\catcode `\$12\catcode
  `\&12\catcode `\#12\catcode `\^12\catcode `\_12\catcode `\%12\relax}%
\providecommand \@@startlink[1]{}%
\providecommand \@@endlink[0]{}%
\providecommand \url  [0]{\begingroup\@sanitize@url \@url }%
\providecommand \@url [1]{\endgroup\@href {#1}{\urlprefix }}%
\providecommand \urlprefix  [0]{URL }%
\providecommand \Eprint [0]{\href }%
\providecommand \doibase [0]{http://dx.doi.org/}%
\providecommand \selectlanguage [0]{\@gobble}%
\providecommand \bibinfo  [0]{\@secondoftwo}%
\providecommand \bibfield  [0]{\@secondoftwo}%
\providecommand \translation [1]{[#1]}%
\providecommand \BibitemOpen [0]{}%
\providecommand \bibitemStop [0]{}%
\providecommand \bibitemNoStop [0]{.\EOS\space}%
\providecommand \EOS [0]{\spacefactor3000\relax}%
\providecommand \BibitemShut  [1]{\csname bibitem#1\endcsname}%
\let\auto@bib@innerbib\@empty
\bibitem [{\citenamefont {Frieman}\ \emph {et~al.}(2008)\citenamefont
  {Frieman}, \citenamefont {Turner},\ and\ \citenamefont
  {Huterer}}]{Frieman:2008sn}%
  \BibitemOpen
  \bibfield  {author} {\bibinfo {author} {\bibfnamefont {J.}~\bibnamefont
  {Frieman}}, \bibinfo {author} {\bibfnamefont {M.}~\bibnamefont {Turner}}, \
  and\ \bibinfo {author} {\bibfnamefont {D.}~\bibnamefont {Huterer}},\ }\href
  {\doibase 10.1146/annurev.astro.46.060407.145243} {\bibfield  {journal}
  {\bibinfo  {journal} {Ann.Rev.Astron.Astrophys.}\ }\textbf {\bibinfo {volume}
  {46}},\ \bibinfo {pages} {385} (\bibinfo {year} {2008})},\ \Eprint
  {http://arxiv.org/abs/0803.0982} {arXiv:0803.0982 [astro-ph]} \BibitemShut
  {NoStop}%
\bibitem [{\citenamefont {Marsh}\ \emph {et~al.}(2014)\citenamefont {Marsh},
  \citenamefont {Bull}, \citenamefont {Ferreira},\ and\ \citenamefont
  {Pontzen}}]{Marsh:2014xoa}%
  \BibitemOpen
  \bibfield  {author} {\bibinfo {author} {\bibfnamefont {D.~J.}\ \bibnamefont
  {Marsh}}, \bibinfo {author} {\bibfnamefont {P.}~\bibnamefont {Bull}},
  \bibinfo {author} {\bibfnamefont {P.~G.}\ \bibnamefont {Ferreira}}, \ and\
  \bibinfo {author} {\bibfnamefont {A.}~\bibnamefont {Pontzen}},\ }\href
  {\doibase 10.1103/PhysRevD.90.105023} {\bibfield  {journal} {\bibinfo
  {journal} {Phys.Rev.}\ }\textbf {\bibinfo {volume} {D90}},\ \bibinfo {pages}
  {105023} (\bibinfo {year} {2014})},\ \Eprint {http://arxiv.org/abs/1406.2301}
  {arXiv:1406.2301 [astro-ph.CO]} \BibitemShut {NoStop}%
\bibitem [{\citenamefont {Ohashi}\ and\ \citenamefont
  {Tsujikawa}(2009)}]{Ohashi:2009xw}%
  \BibitemOpen
  \bibfield  {author} {\bibinfo {author} {\bibfnamefont {J.}~\bibnamefont
  {Ohashi}}\ and\ \bibinfo {author} {\bibfnamefont {S.}~\bibnamefont
  {Tsujikawa}},\ }\href {\doibase 10.1103/PhysRevD.80.103513} {\bibfield
  {journal} {\bibinfo  {journal} {Phys.Rev.}\ }\textbf {\bibinfo {volume}
  {D80}},\ \bibinfo {pages} {103513} (\bibinfo {year} {2009})},\ \Eprint
  {http://arxiv.org/abs/0909.3924} {arXiv:0909.3924 [gr-qc]} \BibitemShut
  {NoStop}%
\bibitem [{\citenamefont {Piazza}\ and\ \citenamefont
  {Tsujikawa}(2004)}]{Piazza:2004df}%
  \BibitemOpen
  \bibfield  {author} {\bibinfo {author} {\bibfnamefont {F.}~\bibnamefont
  {Piazza}}\ and\ \bibinfo {author} {\bibfnamefont {S.}~\bibnamefont
  {Tsujikawa}},\ }\href {\doibase 10.1088/1475-7516/2004/07/004} {\bibfield
  {journal} {\bibinfo  {journal} {JCAP}\ }\textbf {\bibinfo {volume} {0407}},\
  \bibinfo {pages} {004} (\bibinfo {year} {2004})},\ \Eprint
  {http://arxiv.org/abs/hep-th/0405054} {arXiv:hep-th/0405054 [hep-th]}
  \BibitemShut {NoStop}%
\bibitem [{\citenamefont {Tsujikawa}\ and\ \citenamefont
  {Sami}(2004)}]{Tsujikawa:2004dp}%
  \BibitemOpen
  \bibfield  {author} {\bibinfo {author} {\bibfnamefont {S.}~\bibnamefont
  {Tsujikawa}}\ and\ \bibinfo {author} {\bibfnamefont {M.}~\bibnamefont
  {Sami}},\ }\href {\doibase 10.1016/j.physletb.2004.10.023} {\bibfield
  {journal} {\bibinfo  {journal} {Phys.Lett.}\ }\textbf {\bibinfo {volume}
  {B603}},\ \bibinfo {pages} {113} (\bibinfo {year} {2004})},\ \Eprint
  {http://arxiv.org/abs/hep-th/0409212} {arXiv:hep-th/0409212 [hep-th]}
  \BibitemShut {NoStop}%
\bibitem [{\citenamefont {Arkani-Hamed}\ \emph {et~al.}(2004)\citenamefont
  {Arkani-Hamed}, \citenamefont {Cheng}, \citenamefont {Luty},\ and\
  \citenamefont {Mukohyama}}]{ArkaniHamed:2003uy}%
  \BibitemOpen
  \bibfield  {author} {\bibinfo {author} {\bibfnamefont {N.}~\bibnamefont
  {Arkani-Hamed}}, \bibinfo {author} {\bibfnamefont {H.-C.}\ \bibnamefont
  {Cheng}}, \bibinfo {author} {\bibfnamefont {M.~A.}\ \bibnamefont {Luty}}, \
  and\ \bibinfo {author} {\bibfnamefont {S.}~\bibnamefont {Mukohyama}},\ }\href
  {\doibase 10.1088/1126-6708/2004/05/074} {\bibfield  {journal} {\bibinfo
  {journal} {JHEP}\ }\textbf {\bibinfo {volume} {0405}},\ \bibinfo {pages}
  {074} (\bibinfo {year} {2004})},\ \Eprint
  {http://arxiv.org/abs/hep-th/0312099} {arXiv:hep-th/0312099 [hep-th]}
  \BibitemShut {NoStop}%
\bibitem [{\citenamefont {Albrecht}\ \emph {et~al.}(2009)\citenamefont
  {Albrecht}, \citenamefont {Amendola}, \citenamefont {Bernstein},
  \citenamefont {Clowe}, \citenamefont {Eisenstein} \emph
  {et~al.}}]{albrecht:2009ct}%
  \BibitemOpen
  \bibfield  {author} {\bibinfo {author} {\bibfnamefont {A.}~\bibnamefont
  {Albrecht}}, \bibinfo {author} {\bibfnamefont {L.}~\bibnamefont {Amendola}},
  \bibinfo {author} {\bibfnamefont {G.}~\bibnamefont {Bernstein}}, \bibinfo
  {author} {\bibfnamefont {D.}~\bibnamefont {Clowe}}, \bibinfo {author}
  {\bibfnamefont {D.}~\bibnamefont {Eisenstein}},  \emph {et~al.},\ }\href@noop
  {} {\  (\bibinfo {year} {2009})},\ \Eprint {http://arxiv.org/abs/0901.0721}
  {arXiv:0901.0721 [astro-ph.IM]} \BibitemShut {NoStop}%
\bibitem [{\citenamefont {Linder}(2003)}]{Linder:2002et}%
  \BibitemOpen
  \bibfield  {author} {\bibinfo {author} {\bibfnamefont {E.~V.}\ \bibnamefont
  {Linder}},\ }\href {\doibase 10.1103/PhysRevLett.90.091301} {\bibfield
  {journal} {\bibinfo  {journal} {Phys.Rev.Lett.}\ }\textbf {\bibinfo {volume}
  {90}},\ \bibinfo {pages} {091301} (\bibinfo {year} {2003})},\ \Eprint
  {http://arxiv.org/abs/astro-ph/0208512} {arXiv:astro-ph/0208512 [astro-ph]}
  \BibitemShut {NoStop}%
\bibitem [{\citenamefont {Huterer}\ and\ \citenamefont
  {Starkman}(2003)}]{Huterer:2002hy}%
  \BibitemOpen
  \bibfield  {author} {\bibinfo {author} {\bibfnamefont {D.}~\bibnamefont
  {Huterer}}\ and\ \bibinfo {author} {\bibfnamefont {G.}~\bibnamefont
  {Starkman}},\ }\href {\doibase 10.1103/PhysRevLett.90.031301} {\bibfield
  {journal} {\bibinfo  {journal} {Phys.Rev.Lett.}\ }\textbf {\bibinfo {volume}
  {90}},\ \bibinfo {pages} {031301} (\bibinfo {year} {2003})},\ \Eprint
  {http://arxiv.org/abs/astro-ph/0207517} {arXiv:astro-ph/0207517 [astro-ph]}
  \BibitemShut {NoStop}%
\bibitem [{\citenamefont {Huterer}\ and\ \citenamefont
  {Peiris}(2007)}]{Huterer:2006mv}%
  \BibitemOpen
  \bibfield  {author} {\bibinfo {author} {\bibfnamefont {D.}~\bibnamefont
  {Huterer}}\ and\ \bibinfo {author} {\bibfnamefont {H.~V.}\ \bibnamefont
  {Peiris}},\ }\href {\doibase 10.1103/PhysRevD.75.083503} {\bibfield
  {journal} {\bibinfo  {journal} {Phys.Rev.}\ }\textbf {\bibinfo {volume}
  {D75}},\ \bibinfo {pages} {083503} (\bibinfo {year} {2007})},\ \Eprint
  {http://arxiv.org/abs/astro-ph/0610427} {arXiv:astro-ph/0610427 [astro-ph]}
  \BibitemShut {NoStop}%
\end{thebibliography}%

\end{document}